\documentclass[english,american,superscriptaddress,prd,a4paper,nofootinbib,11pt,eqsecnum,secnumarabic]{revtex4-1}
\usepackage{multirow,slashed,float,graphicx,caption,subcaption,hyperref,color}
\usepackage[dvipsnames]{xcolor}
\usepackage[utf8]{inputenc}
\hypersetup{colorlinks=true, linkcolor=BurntOrange, citecolor=BurntOrange,filecolor=BurntOrange, urlcolor=BurntOrange}

\newcommand{\be}{\begin{equation}}
\newcommand{\ee}{\end{equation}}
\newcommand{\bea}{\begin{eqnarray}}
\newcommand{\eea}{\end{eqnarray}}

\def\bsp#1\esp{\begin{split}#1\end{split}}

\usepackage[normalem]{ulem}

\usepackage{epsfig}
\usepackage{amsbsy}
\usepackage{varioref}
\usepackage{pifont}
\usepackage{amsmath,amssymb}
\usepackage{wrapfig}

\usepackage{mdframed}
\usepackage[export]{adjustbox}

\begin{document}

\title{
  Single Vector-Like top quark production via chromomagnetic\\
   interactions at present and future hadron colliders\\
	 A Snowmass 2021 White Paper}

\author{Alexander Belyaev}
\affiliation{School of Physics \& Astronomy, University of Southampton, Southampton SO17 1BJ, UK}
\affiliation{Particle Physics Department, Rutherford Appleton Laboratory, Chilton, Didcot, Oxon OX11 0QX, UK}

\author{R. Sekhar Chivukula}
\affiliation{UC San Diego, 9500 Gilman Drive,  La Jolla, CA 92023-0001, USA}

\author{Benjamin~Fuks}
\affiliation{Laboratoire de Physique Th\'eorique et Hautes Energies (LPTHE), UMR 7589, Sorbonne Universit\'e et CNRS, 4 place Jussieu, 75252 Paris Cedex 05, France}

\author{Elizabeth H. Simmons}
\affiliation{UC San Diego, 9500 Gilman Drive,  La Jolla, CA 92023-0001, USA}

\author{Xing Wang}
\affiliation{UC San Diego, 9500 Gilman Drive,  La Jolla, CA 92023-0001, USA}

\begin{abstract}
In our recent paper~\cite{PhysRevD.104.095024}, we have investigated the potential for the LHC to discover vector-like quark partner states singly produced via their chromomagnetic moment interactions. These production mechanisms extend traditional searches which rely on pair-production of top-quark partner states or on the single production of these states through electroweak interactions, in the sense of providing greatly increased reach in parameter space regions where traditional searches are insensitive.
In this study  we determine the potential of both the 14 TeV high-luminosity LHC (HL-LHC) and a 100 TeV proton-proton collider to probe new vector-like quarks produced in this mode. We focus on  the 
single production of a top-quark partner in association with an ordinary top-quark,
as well as on the resonant production of the bottom-quark partner with its subsequent decay to a top-quark partner and a $W$ boson. For both cases we consider a top-partner decay to the Higgs boson and an ordinary top-quark.
We find that HL-LHC and a future 100 TeV proton collider can probe vector-like partner masses up to about 3 TeV and 15-20 TeV respectively, visibly extending the range of the traditional vector like quark partner searches.
\end{abstract}


\maketitle
\flushbottom

\label{sec:intro}


\section{Introduction}
Theories which provide a dynamical explanation for the large value of the mass of the top quark mass often include TeV-scale vector-like top and bottom partners $T$ and $B$ that can lie in a variety of representations of the electroweak group~\cite{Kaplan:1983fs,Kaplan:1991dc,Agashe:2004rs,Barnard:2013zea,Ferretti:2013kya}. Consequently, these partner states have been actively searched for at the LHC, with analyses dedicated both to their QCD-induced pair production mechanism~\cite{Aaboud:2017zfn,Aaboud:2017qpr,Aaboud:2018xuw,Aaboud:2018saj,Aaboud:2018xpj,Aaboud:2018wxv,Aaboud:2018pii,Sirunyan:2017pks,Sirunyan:2018qau,Sirunyan:2018omb,Sirunyan:2019sza,Sirunyan:2018yun,Sirunyan:2020qvb} and to their electroweak single production mode~\cite{Aaboud:2018saj,Aaboud:2018ifs,Sirunyan:2017ynj,Sirunyan:2018fjh,Sirunyan:2018ncp,Sirunyan:2019xeh,ATLAS:2022ozf}. Whereas the former is motivated by large production cross sections, the latter offers search channels that are relevant when the new quarks are heavy. Current corresponding limits lead to lower bounds on the vector-like quark masses of about 1--1.5~TeV, the exact bounds depending on the assumed vector-like quark decay mode.

Recently, we have estimated the potential of the LHC for an alternative search channel exploiting the often neglected chromomagnetic interactions of the top partners~\cite{PhysRevD.104.095024}. We have showed that for composite scales lying in the TeV-regime, seeking associated signals complements conventional searches for vector-like-quark pair and single production (both for top and bottom partners). In this study we extend our previous analysis to the case of a 100 TeV proton-proton collider (FCC-hh or SppC), and present our findings together with results for the LHC.

We demonstrate that the analysis of the new production modes extend the reach of the LHC to vector-like quarks at run III and after the LHC high-luminosity (HL-LHC) phase in interesting and important regions of the model parameter space. In practice, we focus on an illustrative scenario in which the top-partner subsequently decays to a Higgs boson and an ordinary top quark. 
We find, for the HL-LHC and the FCC-hh/SppC, that top-partner masses ranging up to about 3~TeV  and 20 TeV could be reached respectively, which substantially improves the  expectations for the considered new states. It turns out that this is especially true in regions of parameter space that are inaccessible by traditional searches.

In the next section we briefly describe the simplified-model we use for our investigation, and describe the simulation chain used in our analysis. In Section \ref{sec:spectra} we describe our results, which are summarized in Figs. \ref{figs:final_1} and \ref{figs:final_2} for the HL-LHC and Figs. \ref{figs:100TeV_final_1} and \ref{figs:100TeV_final_2} for FCC-hh/SppC. Finally, a brief ``Executive Summary" is given in Section \ref{sec:executive-summary}.


\section{Model and Simulations}

\label{sec:model}

We consider a simple extension to the Standard Model (SM) with new composite (electroweak) doublets and singlets of vector-like fields directly associated with the generation of the top-quark mass
\begin{equation}
Q^0_{L,R}=
\begin{pmatrix}
	T^0_{L,R} \\ B^0_{L,R}
\end{pmatrix}
\ \ \text{and} \ \ \tilde{T}^0_{L,R}\, .
\end{equation}
These new gauge eigenstates couple to their elementary SM top-quark and bottom-quark counterparts $q_L =(t^0_L, b^0_L)^T$, $t^0_R$, and the SM Higgs doublet $\Phi$ via mass mixings
\begin{equation}
{\cal L}_{\rm mass} =
  -M_Q \overline{Q^0_L} Q^0_R
  -M_{\tilde{T}} \overline{\widetilde{T}^0_L} \widetilde{T}^0_R
  - \Big(
      y^* (\overline{Q^0_L} \cdot \Phi^\dag) \widetilde{T^0_R}
    + \Delta_L \overline{q^0_L} Q^0_R 
    + \Delta_R \overline{t^0_R} \widetilde{T}^0_L
    + {\rm H.c.}
  \Big)\, .
\label{eq:mixing} \end{equation}
In this configuration, we assume that the Higgs field has a composite origin, so we neglect interactions that allows it to directly couple to the elementary quarks (through $\bar{Q}_L\Phi^ct_R$ operators for instance).

The fields introduced above mix into mass eigenstates $t$, $T_1$, $T_2$, $b$ and $B_1$, where $t$ and $b$ stand for the SM botton and top quarks. Moreover, partial compositeness generally predicts the generation of dimension-five chromomagnetic interactions at the electroweak scale,
\be
{\cal L}_{\rm chromo}=\frac{g_s }{\Lambda}\  \overline{{\cal Q}_L} \sigma^{\mu\nu} G_{\mu\nu} {\cal Q}_R + {\rm H.c.}\,  ,
\label{eq:chromomag} \ee
where $\sigma_{\mu\nu} = i(\gamma_\mu\gamma_\nu-\gamma_\nu\gamma_\mu)/2$, and ${\cal Q}_L$ and ${\cal Q}_R$ denote any of the considered left-handed and right-handed new physics gauge eigenstates (${\cal Q} = Q^0$, $\tilde T^0$). The gluon field strength tensor reads $G_{\mu\nu}=G^A_{\mu\nu} T_A$, where the matrices $T_A$ are the fundamental representation matrices of $SU(3)$ (all considered states being color triplets), and we have assumed that the compositeness scale $\Lambda$ is the same for all considered vector-like quarks, which is a natural simplifying assumption. As a result of the mixing of the SM quarks with their composite  partners, the Lagrangian (\ref{eq:chromomag}) gives rise to the  ``off-diagonal" chromomagnetic interactions involving a single third-generation SM quark and a single vector-like quark,
\be\bsp
  {\cal L}_{t} =&\ \frac{g_s}{\Lambda} G_{\mu\nu}  \bigg[
    {\cal C}_1 \overline{T}_{1R} \sigma^{\mu\nu} t_L \!+\! 
    {\cal C}_2 \overline{T}_{1L} \sigma^{\mu\nu} t_R \!+\!
    {\cal C}_3 \overline{T}_{2R} \sigma^{\mu\nu} t_L \!+\!
    {\cal C}_4 \overline{T}_{2L} \sigma^{\mu\nu} t_R \!+\!
    {\cal C}_5 \overline{B}_{1R} \sigma^{\mu\nu} b_L \!+\!
    {\rm H.c.} \bigg]\, , \\
\esp\label{eq:bbg_ttg}\ee
whose couplings ${\cal C}_i$ are functions of the model parameters introduced in Eq.~\ref{eq:mixing} .

These off-diagonal chromomagnetic interactions lead to new single vector-like quark production channels, and open new opportunities for the exploration of vector-like quark physics at colliders, beyond those already accessible and investigated today. The advantage of single production in comparison to the QCD pair production mechanism is the smaller phase space suppression; the advantage of the chromomagnetic-moment-induced single production process relative to electroweak single production is its enhancement due to the strong coupling and a gluon density in the initial state. Moreover, the chromomagnetic operator in Eq.~(\ref{eq:chromomag}) also modifies the ``diagonal'' $gt\bar{t}$ and $gT_1\bar{T}_1$  QCD interactions, and can thus be probed through $t\bar{t}$ and $T_1\bar{T}_1$ pair production~\cite{BuarqueFranzosi:2019dwg}.

The model described above is determined by five independent free mixing parameters and the scale $\Lambda$.  Once we account for the fact that the mass of the SM top quark is known, these parameters can be determined after fixing the following five quantities,
\begin{equation}
\bigg\{ \epsilon_L=\frac{\Delta_L}{M_Q}, \ \ \epsilon_R=\frac{\Delta_R}{M_{\tilde{T}}}, \ \  m_{T_1}, \ \  \frac{M_Q}{M_{\tilde{T}}} \ \ ,  \ \ \Lambda\bigg\}\, , \label{eq:epsilons}
\end{equation}
where $m_{T_1}$ is the mass of the lightest top partner $T_1$.
We focus focus below on two distinct scenarios with $M_Q/M_{\tilde{T}} = 1$ and $2$, and discuss potential search strategies at the LHC and FCC-hh/SppC.

For all results displayed in this paper, we have used {\sc FeynRules} to generate UFO model libraries to be used within the {\sc MadGraph5\_aMC@NLO} package~\cite{Christensen:2009jx,Degrande:2011ua,Alloul:2013bka,Alwall:2014hca}. We then handled the generation of hard-scattering events, that we match with parton showering and hadronisation as modelled by {\sc Pythia}~8~\cite{Sjostrand:2014zea}. Detector simulation has been achieved through {\sc Delphes}~3~\cite{deFavereau:2013fsa} using standard detector parameterizations, and the anti-$k_T$ algorithm~\cite{Cacciari:2008gp} as implemented in {\sc FastJet}~\cite{Cacciari:2011ma} for event reconstruction.

\section{Results}\label{sec:spectra}

\begin{figure}
	\centering
	\begin{subfigure}{0.45\textwidth}
		\includegraphics[width=\textwidth]{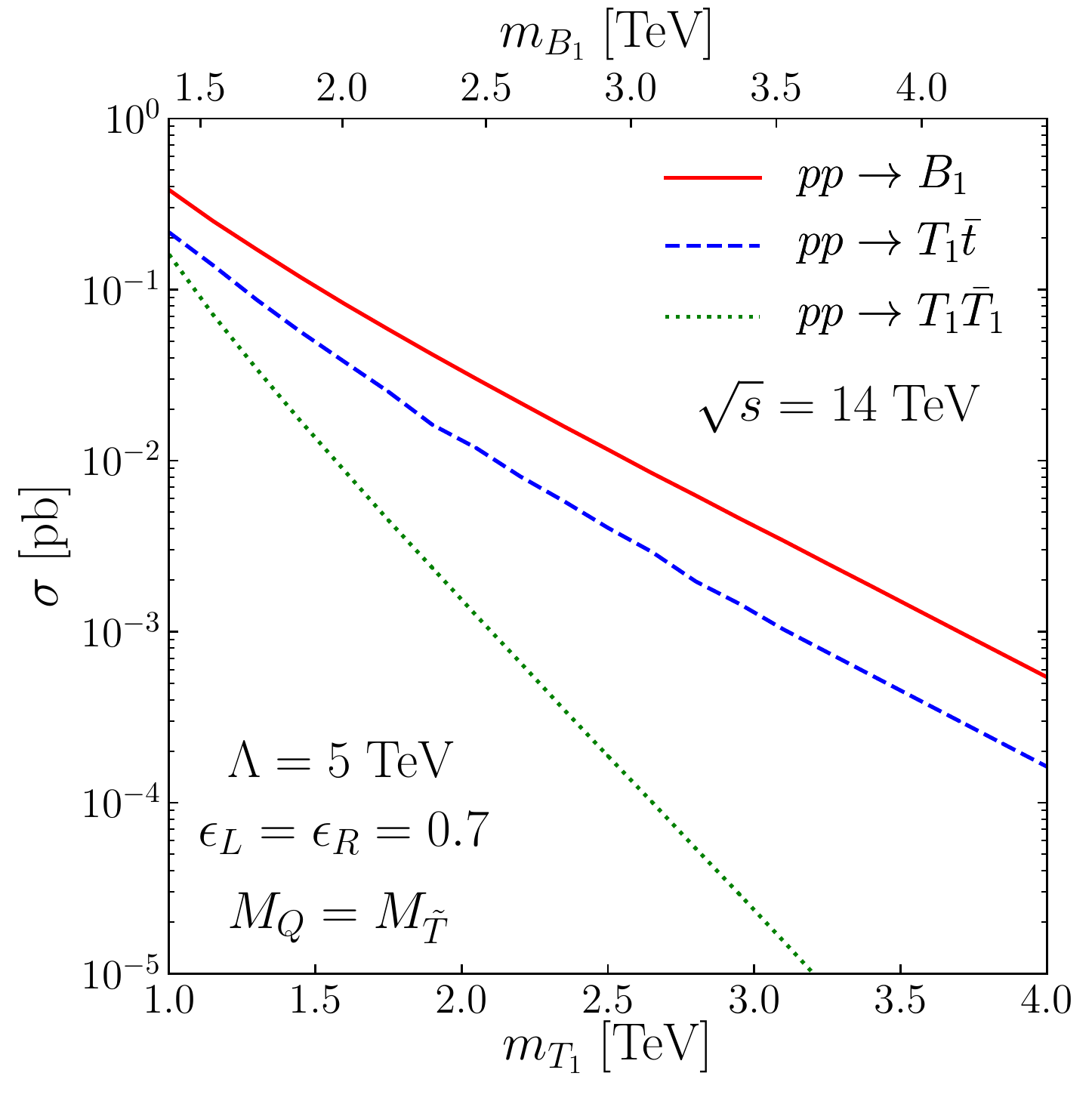}
		\caption{}
	\end{subfigure}
	\begin{subfigure}{0.45\textwidth}
		\includegraphics[width=\textwidth]{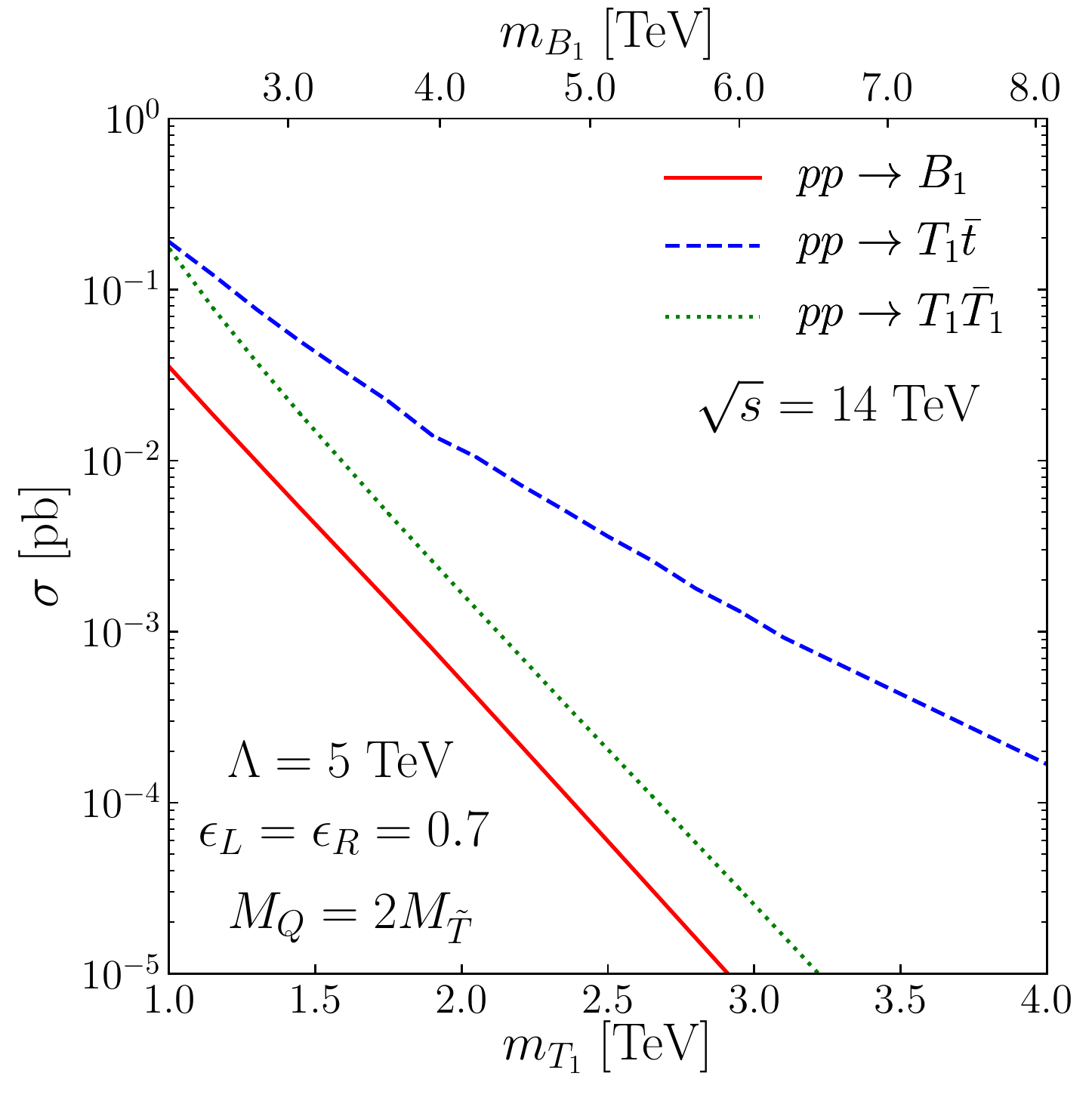}
		\caption{}
	\end{subfigure}
	\\
	\centering
	\begin{subfigure}{0.45\textwidth}
		\includegraphics[width=\textwidth]{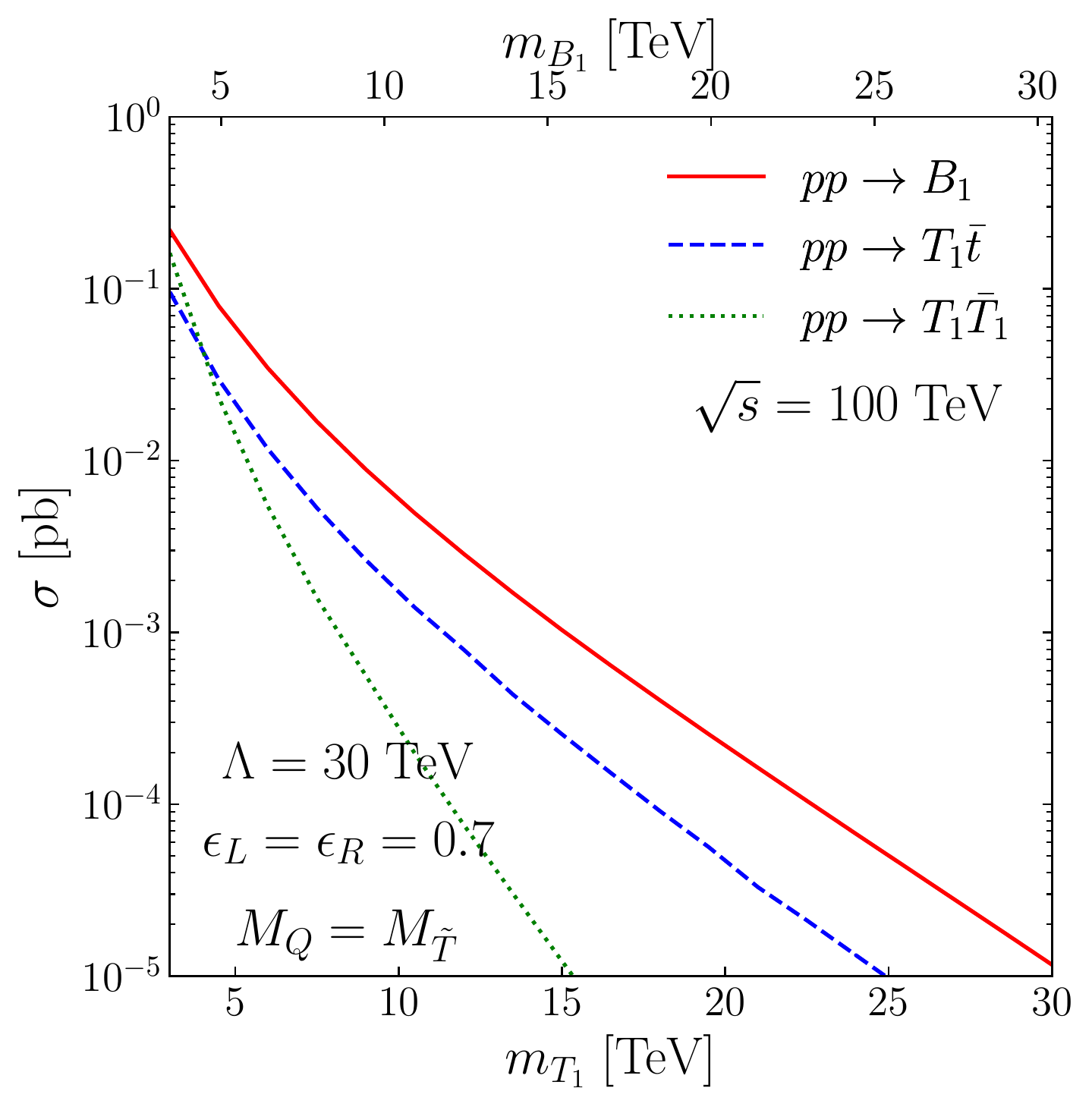}
		\caption{}
	\end{subfigure}
	\begin{subfigure}{0.45\textwidth}
		\includegraphics[width=\textwidth]{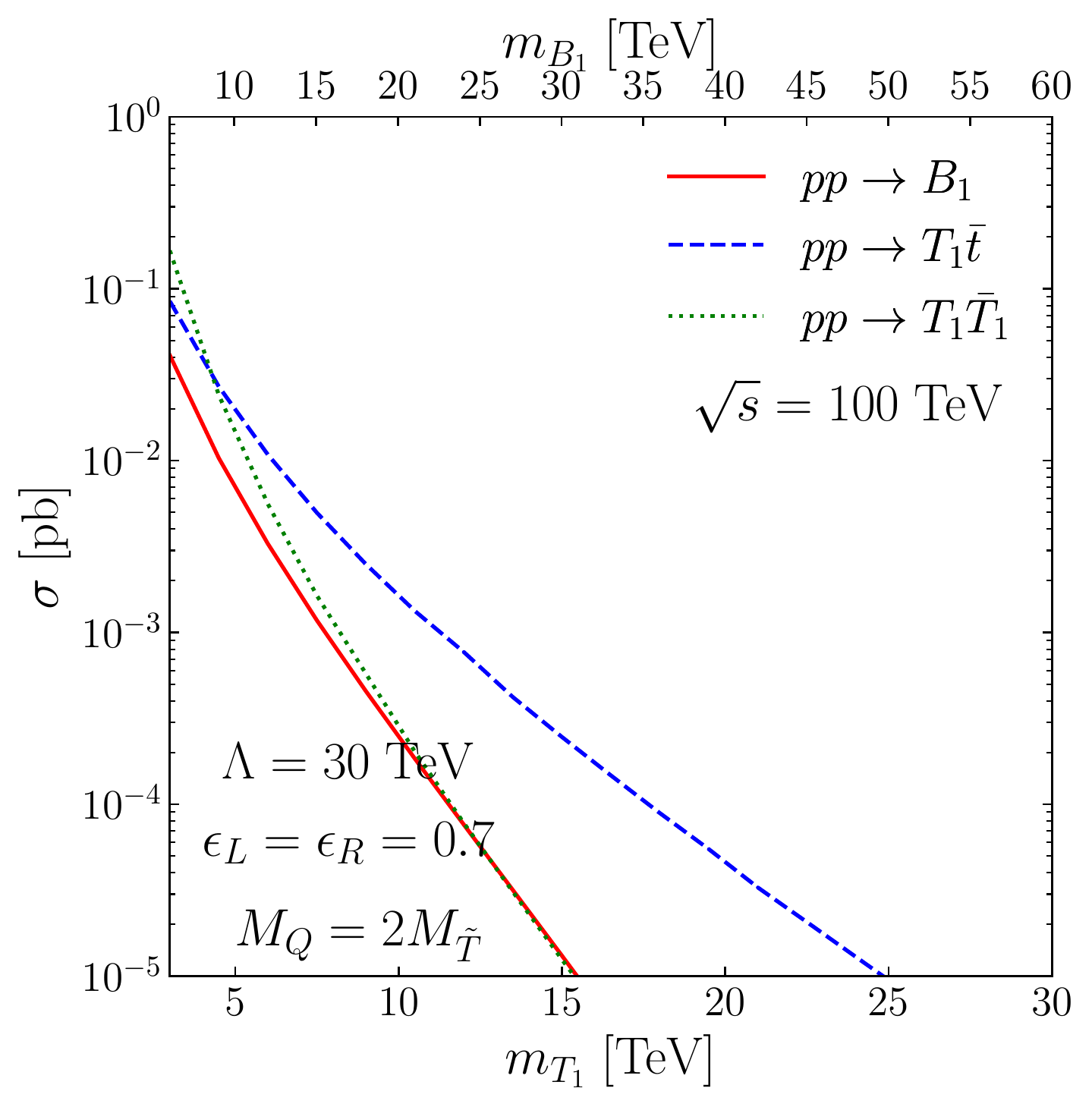}
		\caption{}
	\end{subfigure}

	\caption{Production cross sections for the processes $pp\to B_1$ (solid red), $pp\to T_1\bar t$ (dashed blue) and $pp\to T_1\bar T_1$ (dotted green) at $\sqrt{s}=14$~TeV (top row) and $\sqrt{s}=100$~TeV (bottom row). The first two processes occur through the chromomagnetic moment operators, whereas the last process receives contributions from both pure QCD processes and the chromomagnetic moment operators. We consider as a benchmark scenario a configuration in which $\epsilon_L = \epsilon_R = 0.7$, $\Lambda = 5$ TeV for $\sqrt{s} = 14$ TeV ($\Lambda = 30$ TeV for $\sqrt{s} = 100$ TeV), and in which (a)  $M_Q = M_{\tilde{T}}$ (left) or (b) $M_Q = 2 M_{\tilde{T}}$ (right).} \label{figs:xsec}
\end{figure}

The chromomagnetic operators in Eq. (\ref{eq:bbg_ttg}) give rise to two novel vector-like quark single-production mechanisms via the partonic processes $bg \to B_1$ (resonant production) and $(q\bar{q}, gg) \rightarrow T_1 t$ (via $s$-channel gluon exchanges). We summarize the expected rates these production mechanisms and compare them to the pair-production ones in Fig.~\ref{figs:xsec}, the rate being shown as a function of the mass of the $T_1$ state $m_{T_1}$. The corresponding $B_1$ mass is shown on the upper axis in each figure. Note that the single production rates through the chromomagnetic operators can substantially exceed the rates for pair-production at high veector-like quark mass.

In the case of $M_Q = M_{\tilde{T}}$, the top and bottom partners $T_1$ and $B_1$ have similar masses and $B_1$ single production is the dominant production channel. The corresponding cross section is a factor of a few larger than that corresponding to the $pp\to T_1\bar t$ process, as shown in Fig.~\ref{figs:xsec}(a). While single $B_1$ production has been searched through the conventional decay channels at the LHC, $B_1\rightarrow t W$ and $B_1\rightarrow g g$~\cite{CMS:2021ptb,CMS:2019gwf}, the sensitivity becomes much weaker for scenarios in which $B_1$ dominantly decays into a $T_1W$ system, a configuration that is realized when $\epsilon_L > \epsilon_R$. When $M_Q = 2M_{\tilde{T}}$, the bottom partner $B_1$ is much heavier than the $T_1$ state, so that the $2\to 1$ process exhibits a significant phase-space suppression; the dominant vector-like quark production mode then becomes $T_1t$ associated production.

\subsection{Single Production of the $B_1$ Partner}\label{sec:ppB}

In our previous study \cite{PhysRevD.104.095024}, we focused on the complementary search for the decay mode $B_1 \rightarrow T_1 W$, the heavy top partner then undergoing a subsequent decay into a (boosted) top quark and a (boosted) Higgs boson that itself decays into a pair of $b$-jets,
\begin{equation}
	pp\to B_1\to T_1 W \to (t\ h) W \qquad\text{with}\qquad h\rightarrow b\bar{b}\, .
	\label{eq:Bsignal}\end{equation}
The choice of $T_1 \to th$ decay channel for our study is twofold. On one hand this decay channel is either dominant or 
non-negligible in the whole parameter space~\cite{PhysRevD.104.095024}. On the other hand,
such a final state features a high $b$-jet multiplicity which 
together with the boosted heavy Higgs and top ``jets" can be used to efficiently suppress the SM background. In order to avoid dealing with the rejection of the overwhelming QCD multi-jet background, we consider a signature in which the top quark decays hadronically while the $W$ boson originating from the initial $B_1$ decay decays leptonically. In this case, the dominant contributions to the SM background originate from $t\bar{t}$ production (in association with jets), which we will take into account to estimate the collider sensitivity to the signal (\ref{eq:Bsignal}). Any other potential contribution to the background is neglected, and is thus assumed to become negligbly small after the analysis selection.

The final-state boosted top quark and Higgs boson are reconstructed as $R=0.8$ fat jets, that we identify through their soft-drop mass $M_{\rm SD}$ and which we require to feature a substructure with one and two $R=0.2$ slim $b$-tagged jet, respectively. We optimize the significance by using a boosted decision tree (BDT)~\cite{Friedman:2001wbq}, with the input variables given by\footnote{More details about event reconstruction and event selection can be found in  \cite{PhysRevD.104.095024}.}
\be
\left\{\bsp
&p_T(t_{\rm had}),\ p_T(W_{\rm lep}),\ p_T(h),\
p_T(b_{\rm had}),\ p_T(b_{h1}),\ p_T(b_{h2}),\
\eta(t_{\rm had}),\ \eta(W_{\rm lep}),\ \eta_T(h),\\
&\Delta R(t_{\rm had}, W_{\rm lep}),\  \Delta R(t_{\rm had}, h),\ \Delta R(W_{\rm lep}, h),\
\Delta R(t_{\rm had}, b_{\rm had}),\\
&m(t_{\rm had}, W_{\rm lep}),\  m(t_{\rm had}, h),\ m(W_{\rm lep}, h),
m(t_{\rm had}, W_{\rm lep}, h),\
S_T,\ S_{T, {\rm reco}}
\esp\right\}\, .
\ee
This set of variable spans a variety of transverse momenta and pseudorapidities, angular distances and invariant masses of various systems of reconstructed objets.

\subsection{Single Production of the $T_1$ Partner}\label{sec:tT1}

When the masses of the lighter $T_1$ and heavier $B_1$ states are quite split, vector-like quark production proceeds dominantly through the associated production of a $T_1$ partner with a top quark. We consider a $T_1$ decay into a boosted Higgs boson and a (boosted) top quark, and we then focus again on a Higgs boson decaying into a $b\bar b$ system. The full process thus reads
\begin{equation}
	p p \to T_1 t \rightarrow (t\ h)\ t \qquad\text{with}\qquad h\rightarrow b\bar{b}\, .
\end{equation}
Once again, such a process features a high $b$-jet multiplicity, that we could use as a handle for background rejection. In order to evade the overwhelming QCD multi-jet background, we consider a signal topology in which the boosted top quark ({\it i.e.}\ the top quark that originates from the top-partner decay) decays hadronically while the spectator top quark ({\it i.e.}\ the top quark that is produced in association with the $T_1$ state) decays semi-leptonically. As in the case of the previous signal, the dominant contributions to the SM background arise from $t\bar{t}+$jets production, which we consider as the sole background in our analysis. Any other potential background components are indeed expected to be subleading after the selection cuts of our analysis, and are therefore neglected.

The signal significance is optimized through a BDT, with the input variables including transverse momenta, pseudorapidities, angular separations and invariant masses of the reconstructed objects,\footnote{Again, details can be found in  \cite{PhysRevD.104.095024}.}
\be\left\{\bsp
&p_T(t_{\rm had}),\ p_T(t_{\rm lep}),\ p_T(h),\
p_T(b_{\rm had}),\ p_T(b_{\rm lep}),\ p_T(b_{h1}),\ p_T(b_{h2}),\\
&\eta(t_{\rm had}),\ \eta(t_{\rm lep}),\ \eta_T(h),\\
&\Delta R(t_{\rm had}, t_{\rm lep}),\  \Delta R(t_{\rm had}, h),\ \Delta R(t_{\rm lep}, h),\
\Delta R(t_{\rm had}, b_{\rm had}),\\
&m(t_{\rm had}, t_{\rm lep}),\  m(t_{\rm had}, h),\ m(t_{\rm lep}, h),
m(b_{\rm lep}, \ell, \nu),\
m(t_{\rm had}, t_{\rm lep}, h),\
S_T,\ S_{T, {\rm reco}}
\esp\right\}\, .\ee

\subsection{Projected sensitivity at HL-LHC}\label{sec:hllhc}
\begin{figure}[h]
\centering
\begin{subfigure}{0.33\textwidth}
\includegraphics[width=\textwidth]{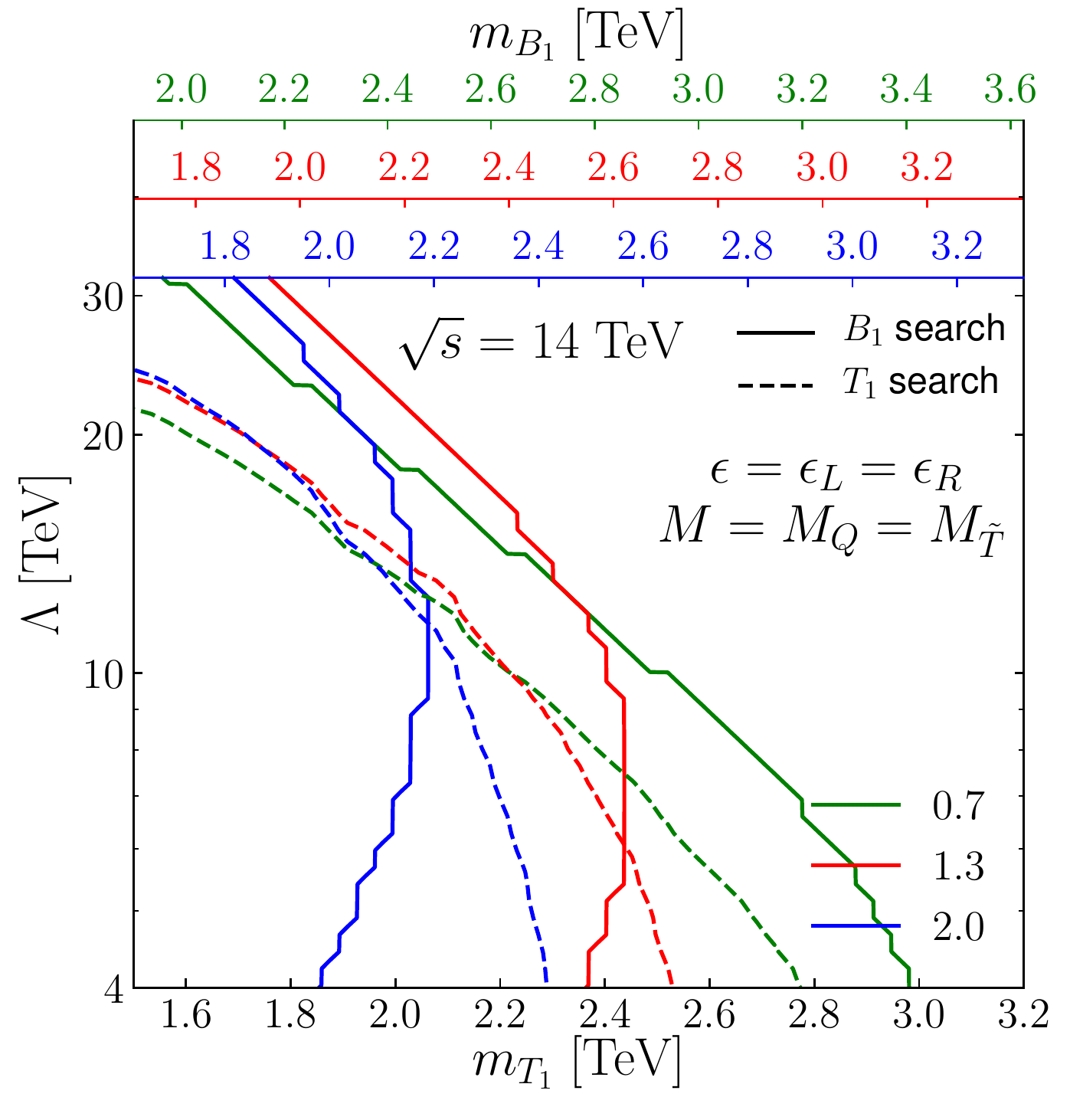}
\caption{}
\end{subfigure}%
\begin{subfigure}{0.33\textwidth}
\includegraphics[width=\textwidth]{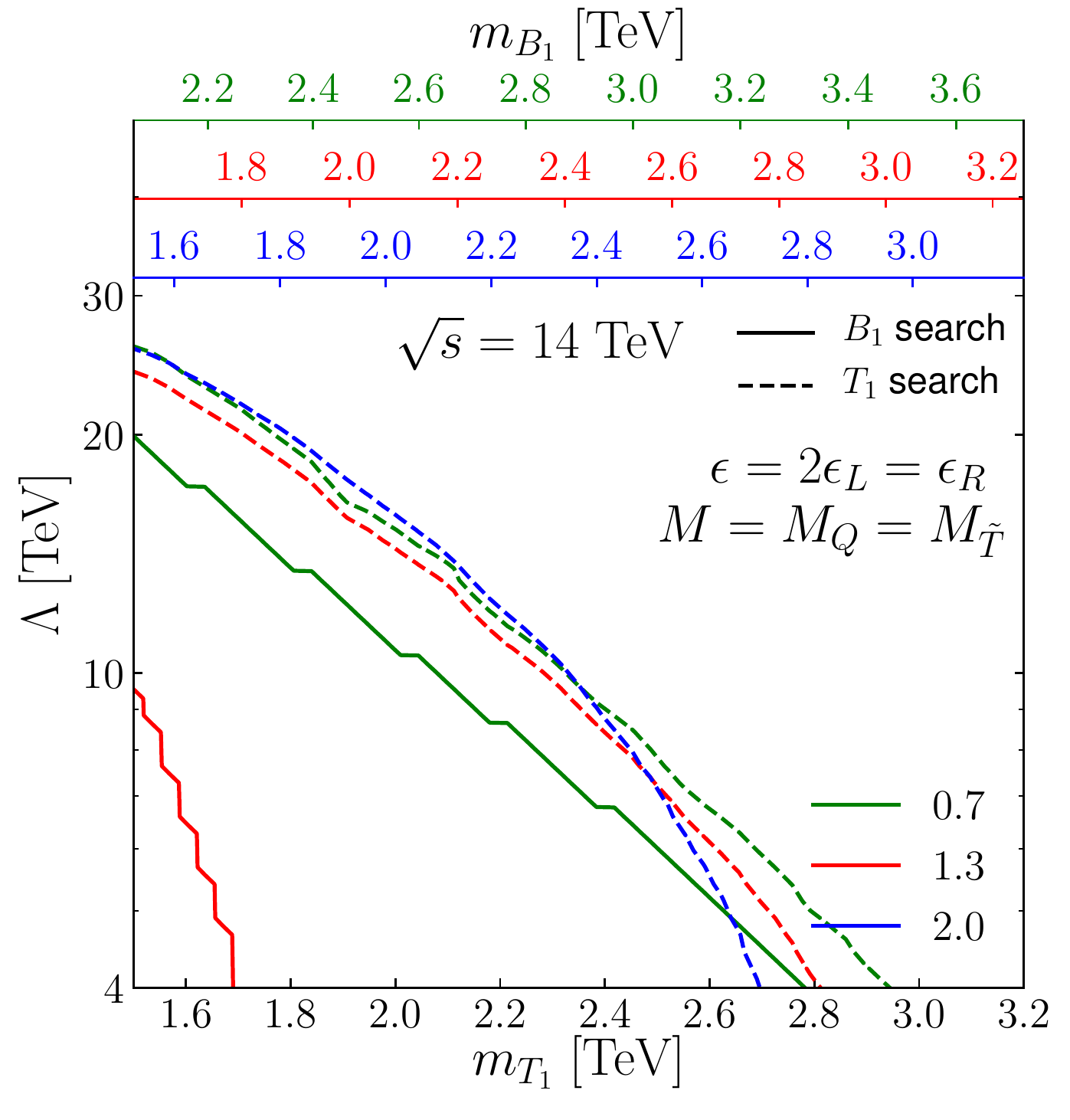}
\caption{}
\end{subfigure}
\begin{subfigure}{0.33\textwidth}
\includegraphics[width=\textwidth]{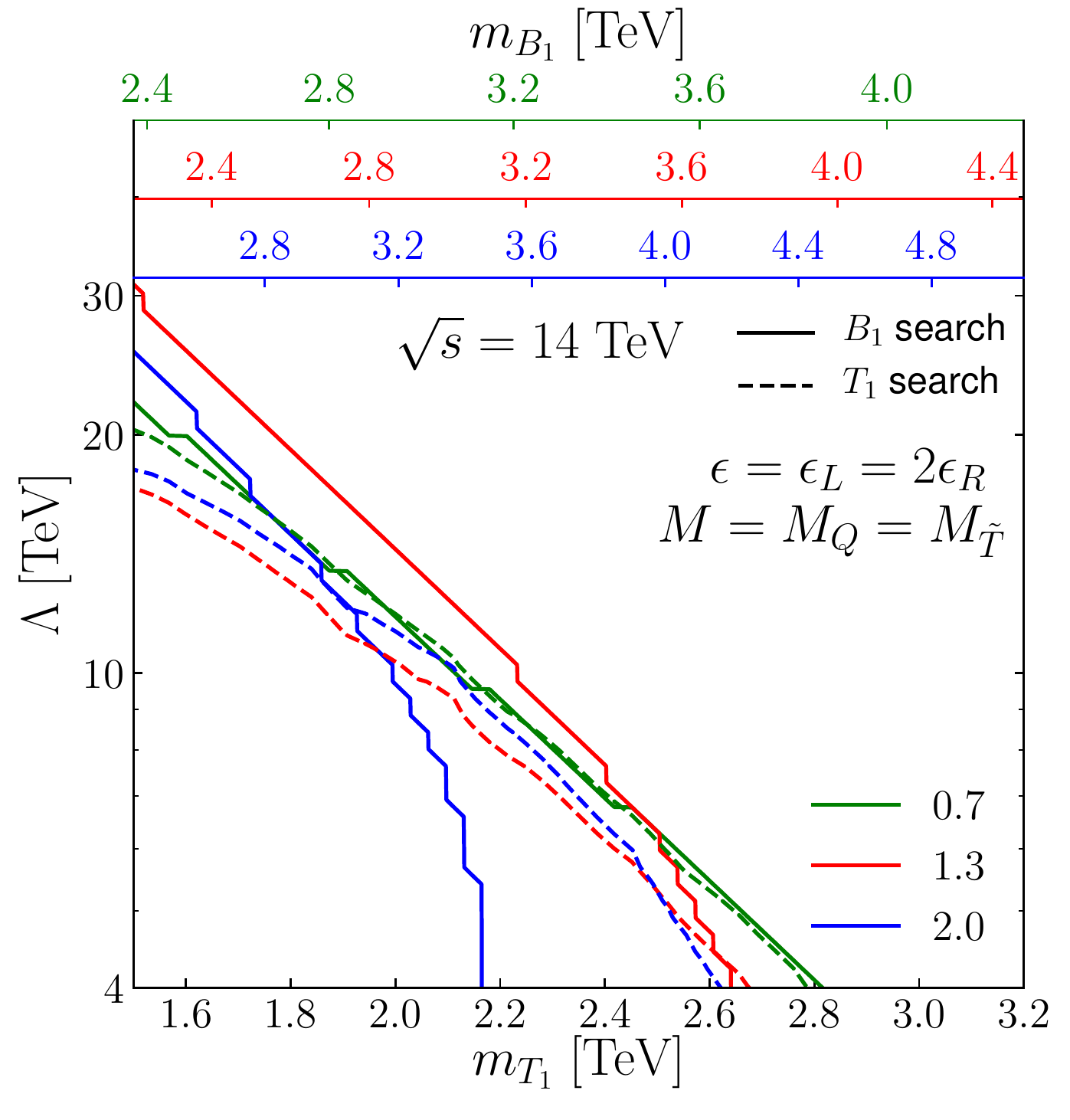}
\caption{}
\end{subfigure}
\caption{HL-LHC sensitivity to the $pp\rightarrow B_1\rightarrow T_1W$ (solid) and $pp\rightarrow T_1 t$ (dashed) signals, shown as two-dimensional 95\% C.L.\ exclusion contours in the $(m_{T_1}, \Lambda)$ plane, with the excluded regions being toward the lower left in each diagram. We  study scenarios in which $M_Q=M_{\tilde{T}}$ and $\epsilon = 0.7$ (green), $1.3$ (red), and $2$ (blue), the mixing being defined as $\epsilon \equiv \epsilon_L = \epsilon_R$ (a), $\epsilon \equiv 2\epsilon_L = \epsilon_R$ (b) and $\epsilon \equiv \epsilon_L = 2\epsilon_R$ (c). The corresponding $B_1$-quark masses $m_{B_1}$ are shown at the top of the figures.}
\label{figs:final_1}
\end{figure}

\begin{figure}
	\centering
	\begin{subfigure}{0.33\textwidth}
		\includegraphics[width=\textwidth]{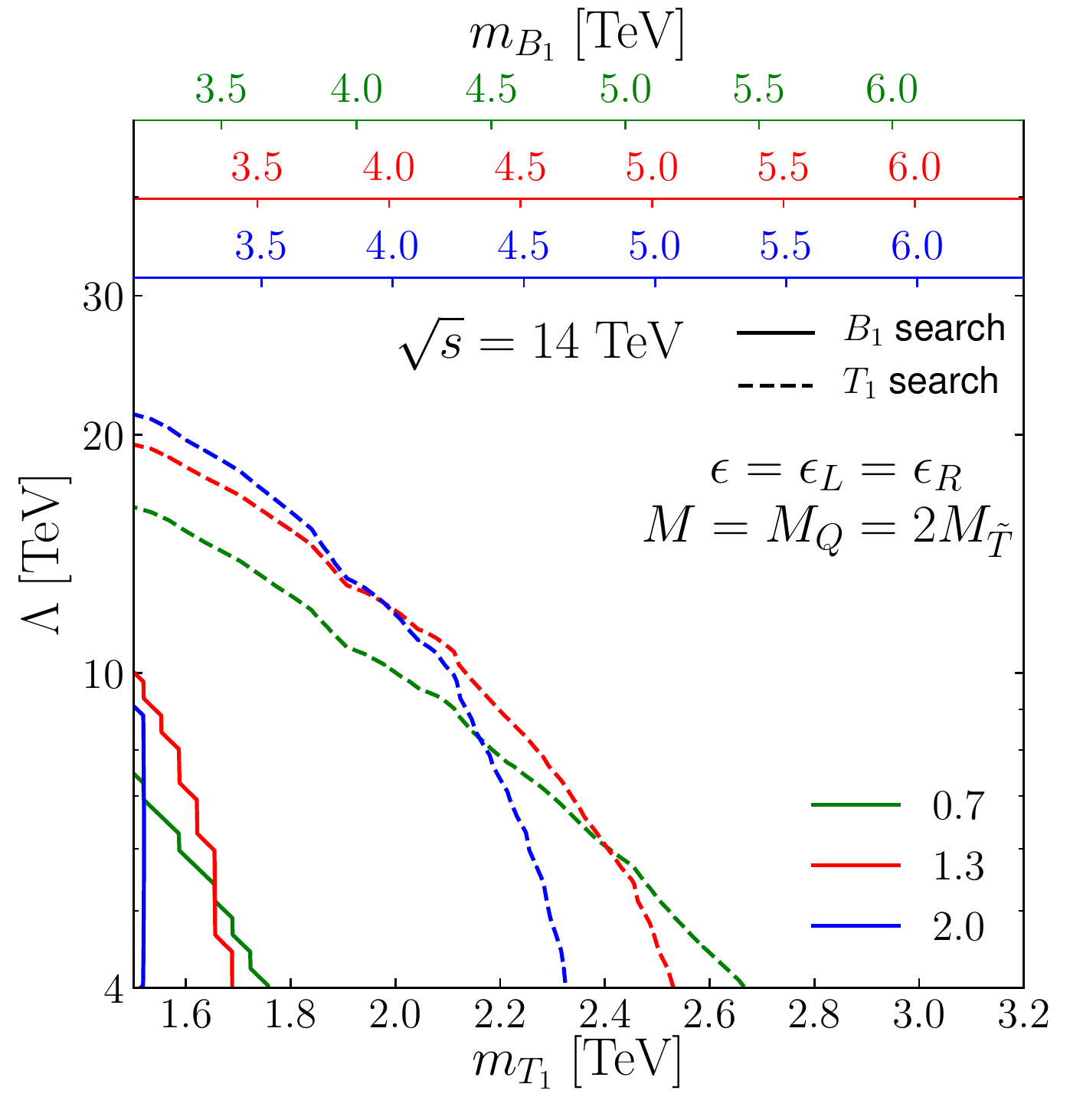}
		\caption{}
	\end{subfigure}%
	\begin{subfigure}{0.33\textwidth}
		\centering
		\includegraphics[width=\textwidth]{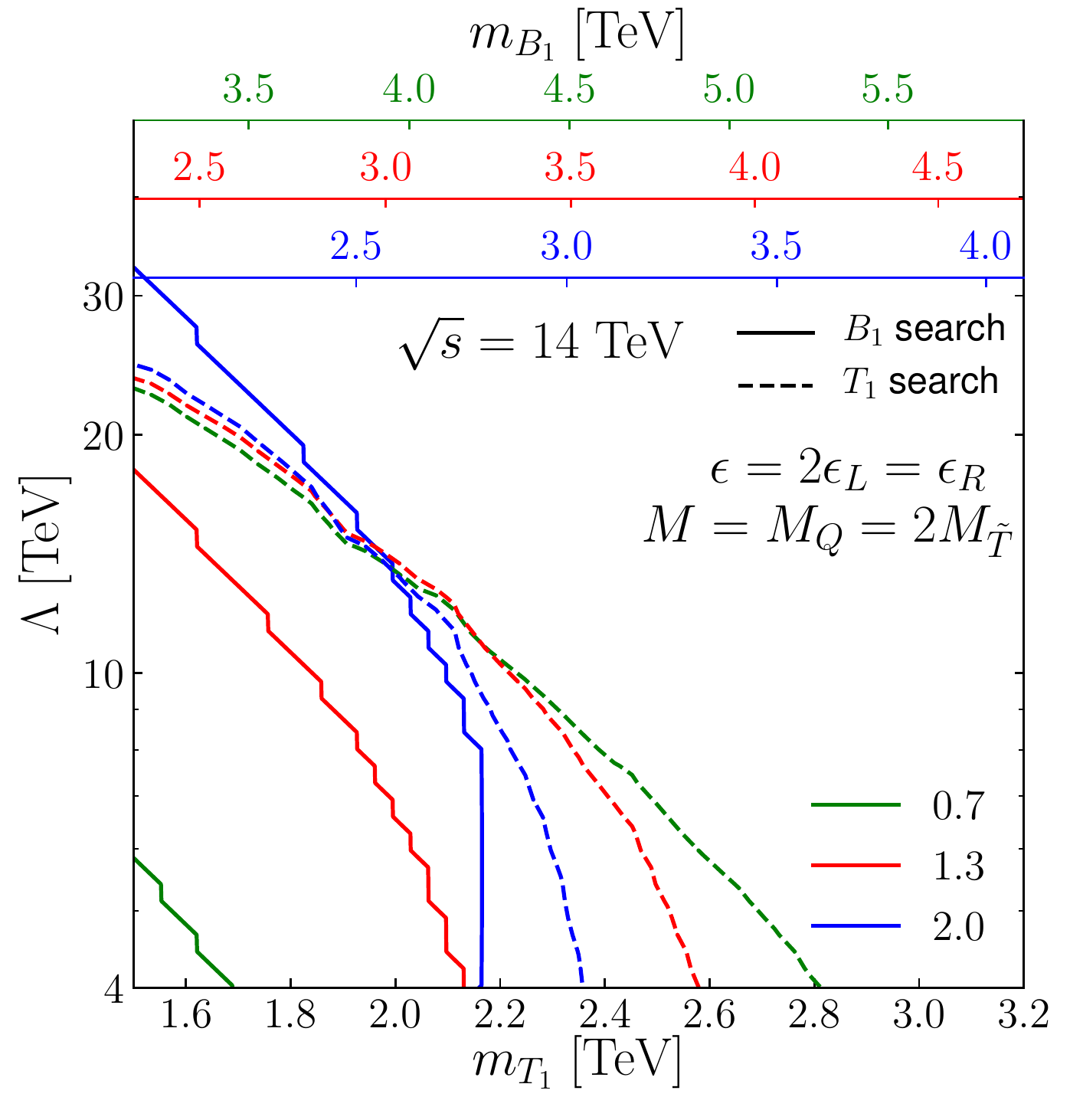}
		\caption{}
	\end{subfigure}
	\begin{subfigure}{0.33\textwidth}
		\centering
		\includegraphics[width=\textwidth]{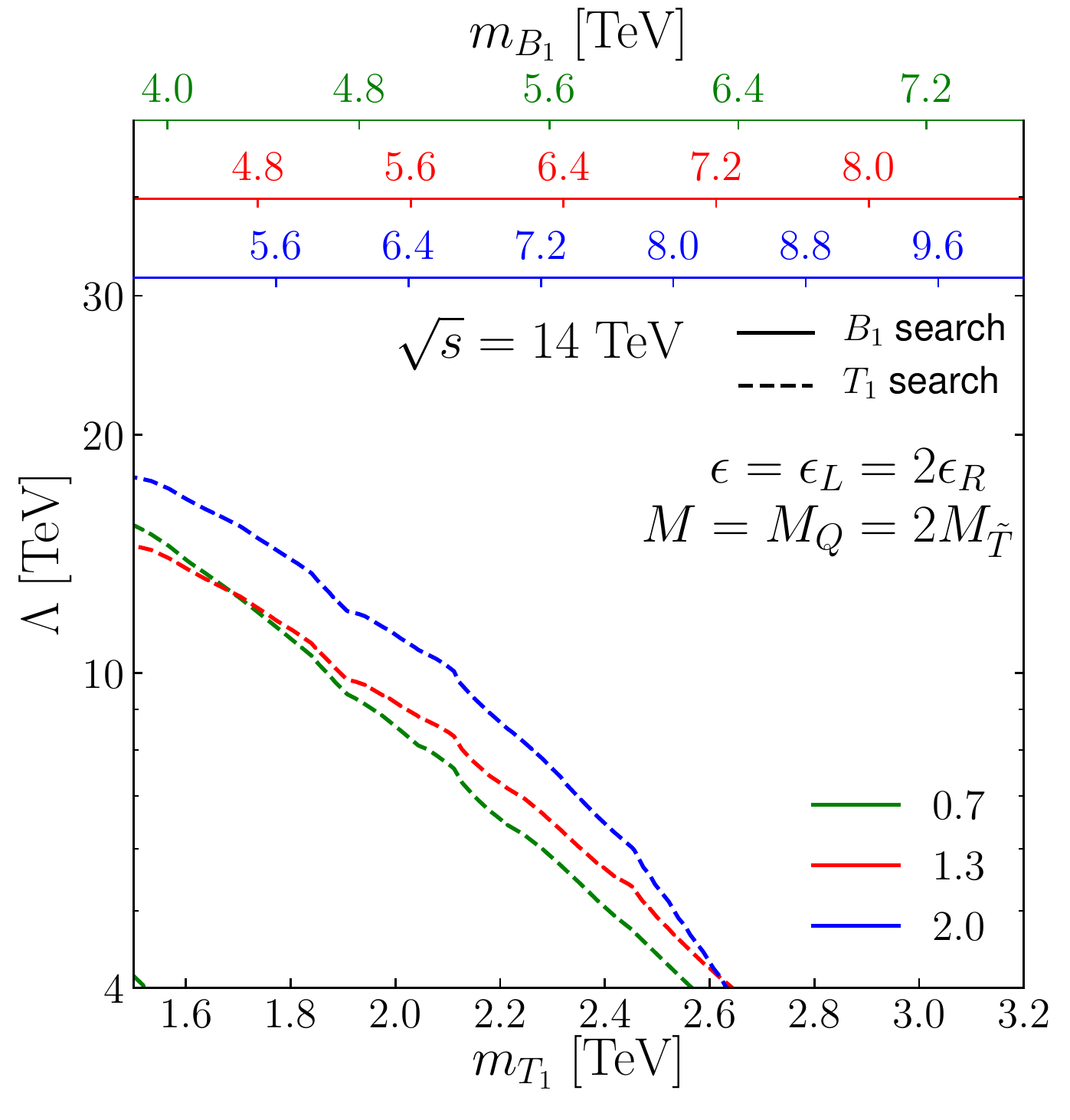}
		\caption{}
	\end{subfigure}
	\caption{Same as Fig.~\ref{figs:final_1} but for $M_Q=2 M_{\tilde{T}}$.}
	\label{figs:final_2}
\end{figure}

To estimate quantitatively the sensitivity of the HL-LHC (and, as discussed in the next section,  the FCC-hh/SppC) to the two signals considered, we define their statistical significance $Z$~\cite{Cousins:2008zz,Cowan:2010js}
\begin{equation}
	Z = \sqrt{2\left((S+B)\ln\frac{S+B}{B} - S\right)}\, ,
\end{equation}
where $S$ and $B$ are the numbers of events for signal and background after all selection cuts respectively. The selection includes a cut on the BDT scores, which has been chosen optimally. Moreover, we require that at least 3 signal events survive the selection ({\it i.e.}\ $S\geq 3$). The resulting two-dimensional 95\% confidence level (C.L.) contours are presented in the $(m_{T_1}, \Lambda)$ plane in Figs.~\ref{figs:final_1} and \ref{figs:final_2} for various scenarios (the excluded regions are at the lower left in each diagram).

In Fig.~\ref{figs:final_1}, we choose that $M_Q = M_{\tilde T}$ whereas in Fig.~\ref{figs:final_2}, we have $M_Q = 2 M_{\tilde T}$. The mixing parameters are taken such that (a) $\epsilon \equiv \epsilon_L = \epsilon_R$, (b) $\epsilon \equiv 2\epsilon_L = \epsilon_R$, or (c) $\epsilon \equiv \epsilon_L = 2\epsilon_R$, and we examine configurations in which the mixing is small ($\epsilon=0.7$; green), moderate ($\epsilon=1.3$; red) and larger ($\epsilon=2$; blue). The exclusions associated with the $pp\to B_1$ analysis (section~\ref{sec:ppB}) are then shown as solid lines, and those associated with the $pp\to T_1t$ analysis (section~\ref{sec:tT1}) are depicted through dashed lines. In each figure, we additionally include an upper horizontal axis for each studied $\epsilon$ value, that we use to represent the corresponding $B_1$ mass values.

For $M_Q = M_{\tilde T}$ (Fig.~\ref{figs:final_1}), the largest obtained sensitivity for the parameter range studied in terms of heavy quark masses corresponds to scenarios in which $\epsilon = 0.7$ and $\Lambda = 4$~TeV.\footnote{We do not explore values of $\Lambda$ less than 4.0 TeV in order to maintain the reliability of the effective field theory considered for the partner masses of interest.} In these cases, bottom-quark partners with masses ranging up to $m_{B_1}\in [3.2, 3.6]$~TeV can be probed. Very importantly, the existing searches are especially insensitive to scenarios in which $\epsilon = \epsilon_L = 2\epsilon_R$, due to suppressed branching fractions of $B_1\rightarrow tW, gg$. Our results show that there is actually a quite promising LHC sensitivity to this configuration, allowing for discovery of both top- and bottom-partners provided extra channels such as those proposed in the present study are added to the LHC experimental program. Compositeness scales of $\Lambda = 20-30$~TeV can even be reached for top-partner masses of about 1.5~TeV.

Once the spectrum becomes more split and the $m_{B_1}$ mass is much larger than the $m_{T_1}$ mass, as shown in Fig.~\ref{figs:final_2} where $M_Q = 2 M_{\tilde T}$, the direct production process $pp\to B_1$ is suppressed, and the expected LHC reach for the models is therefore entirely dictated by the search for $pp\to T_1t$. The largest obtained sensitivity in terms of heavy quark masses corresponds again to the smallest $\Lambda$ values considered ({\it i.e.}\ $\Lambda= 4$~TeV), top-quark partners with masses ranging up to about $[2.4, 2.8]$~TeV being reachable regardless of the heavy quark mixing parameters. In this configuration, existing searches have no sensitivity (regardless of the relative sizes of $\epsilon_L$ and $\epsilon_R$), so that the analysis proposed in Section~\ref{sec:tT1} provides a novel (and unique, so far) promising avenue to explore realistic composite models at the LHC.

\subsection{Projected sensitivity at FCC-hh/SppC}

\begin{figure}[t]
\centering
\begin{subfigure}{0.33\textwidth}
\includegraphics[width=\textwidth]{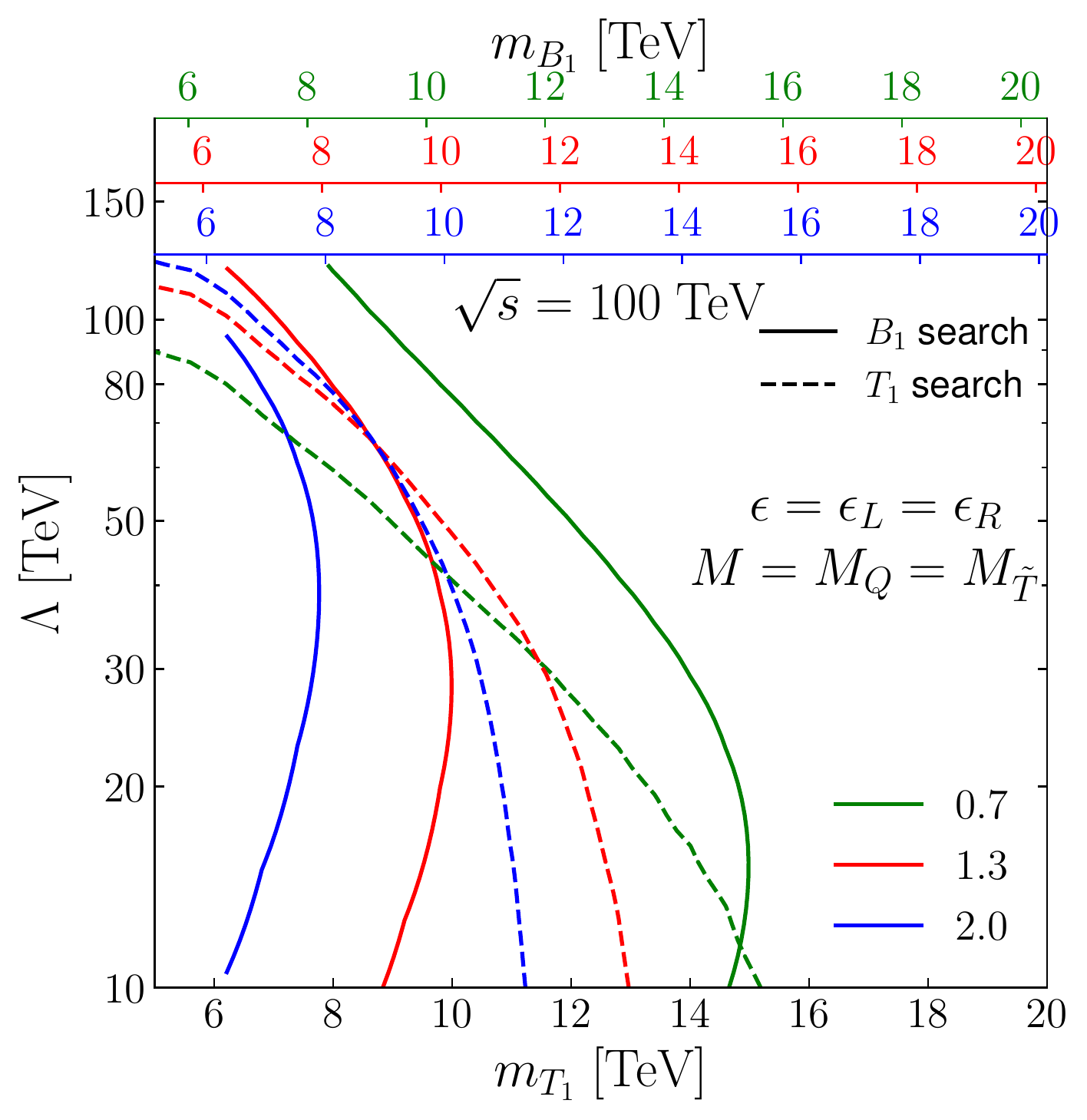}
\caption{}
\end{subfigure}%
\begin{subfigure}{0.33\textwidth}
\includegraphics[width=\textwidth]{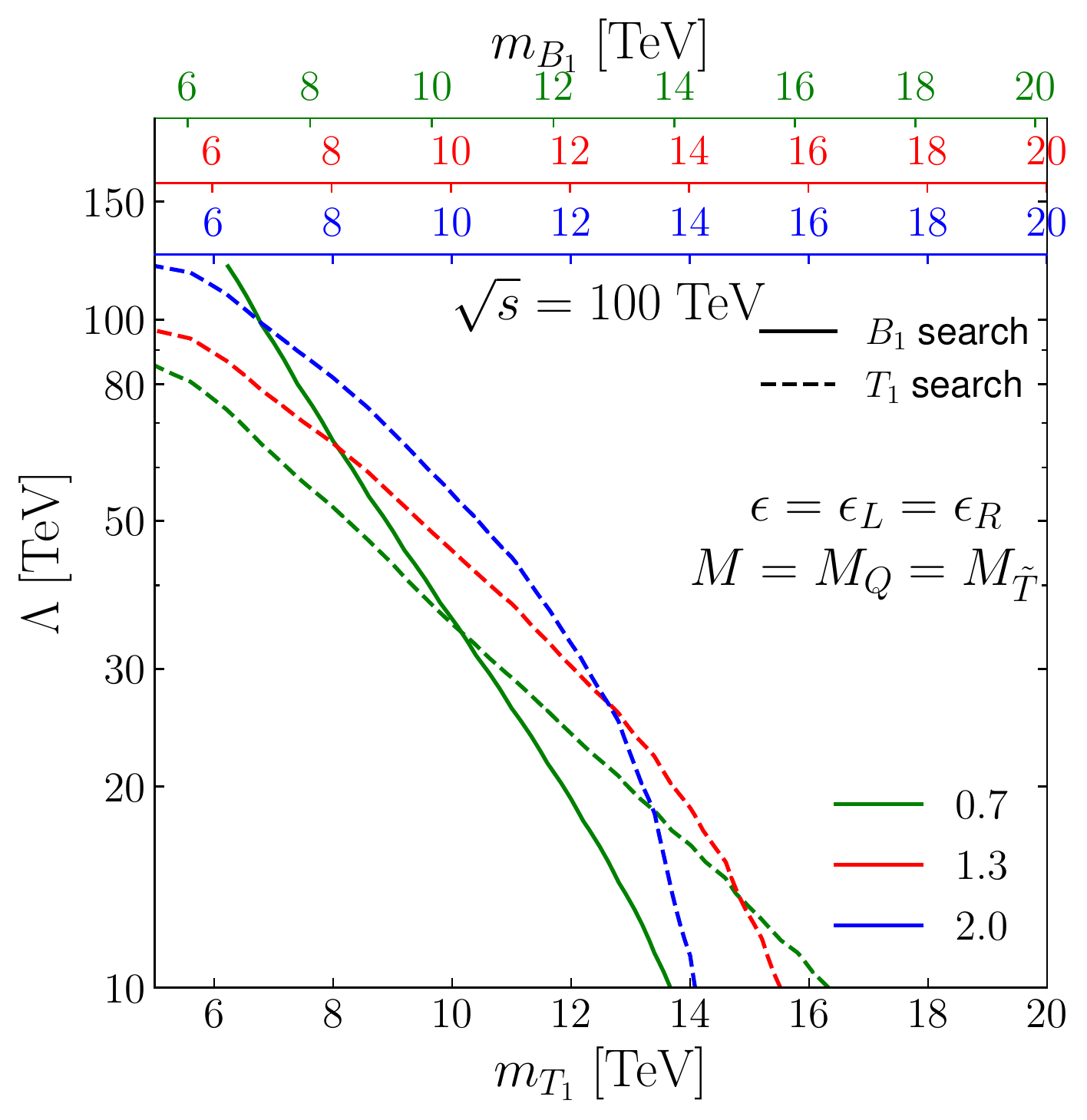}
\caption{}
\end{subfigure}
\begin{subfigure}{0.33\textwidth}
\includegraphics[width=\textwidth]{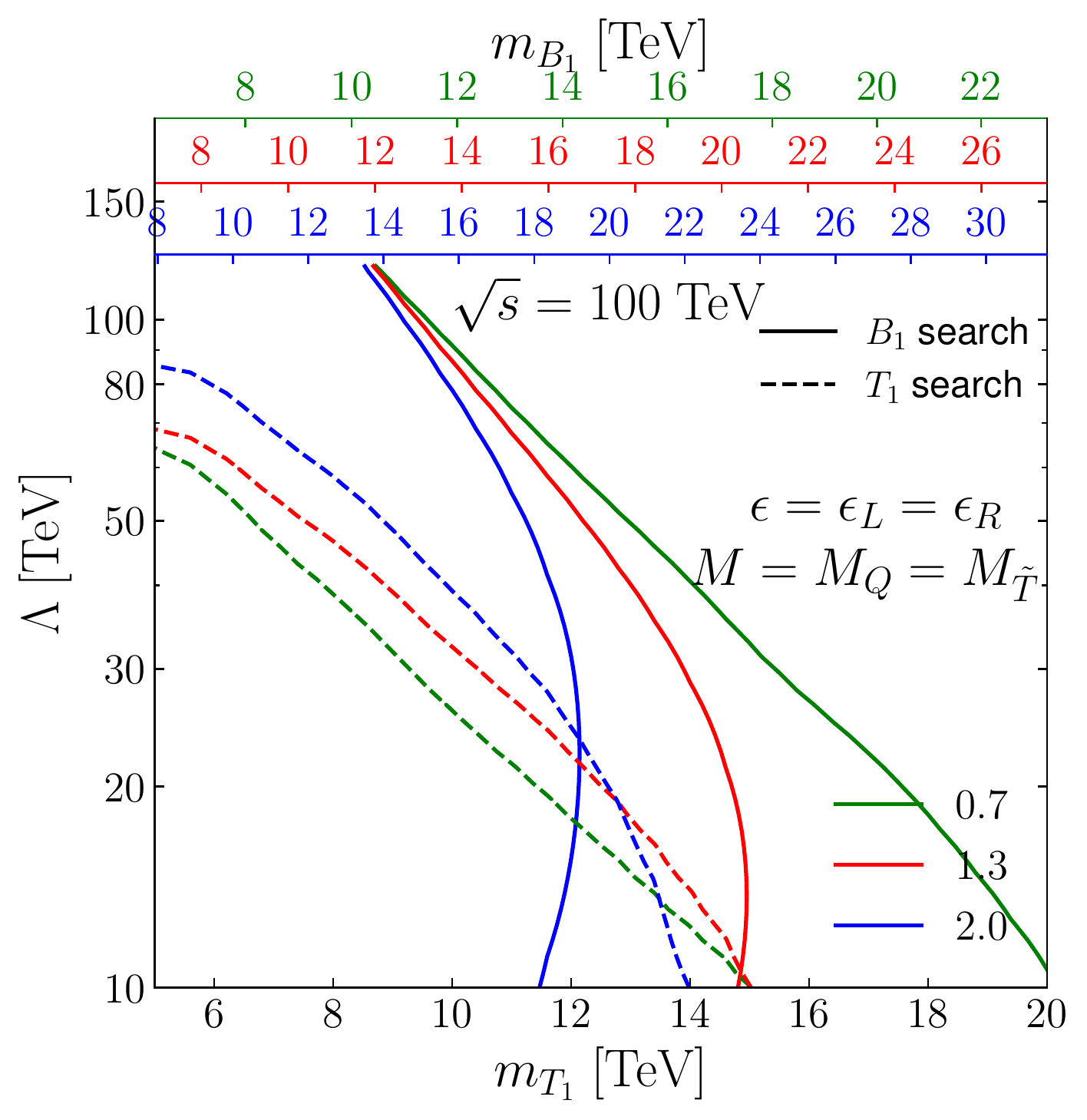}
\caption{}
\end{subfigure}
\caption{HL-LHC sensitivity to the $pp\rightarrow B_1\rightarrow T_1W$ (solid) and $pp\rightarrow T_1 t$ (dashed) signals, shown as two-dimensional 95\% C.L.\ exclusion contours in the $(m_{T_1}, \Lambda)$ plane, with the excluded regions being toward the lower left in each diagram. We  study scenarios in which $M_Q=M_{\tilde{T}}$ and $\epsilon = 0.7$ (green), $1.3$ (red), and $2$ (blue), the mixing being defined as $\epsilon \equiv \epsilon_L = \epsilon_R$ (a), $\epsilon \equiv 2\epsilon_L = \epsilon_R$ (b) and $\epsilon \equiv \epsilon_L = 2\epsilon_R$ (c). The corresponding $B_1$-quark masses $m_{B_1}$ are shown at the top of the figures.}
\label{figs:100TeV_final_1}
	\begin{subfigure}{0.33\textwidth}
		\includegraphics[width=\textwidth]{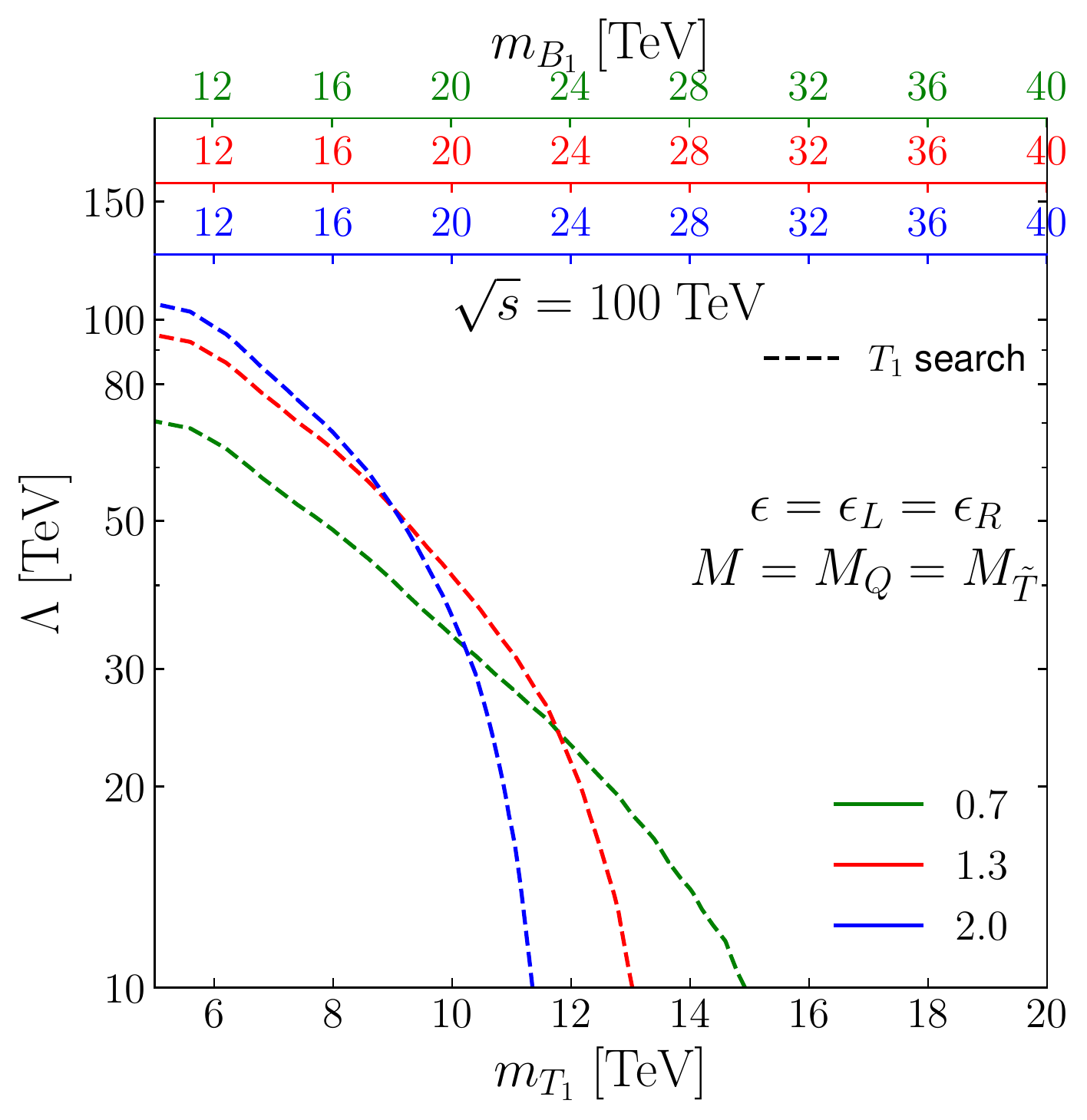}
		\caption{}
	\end{subfigure}%
	\begin{subfigure}{0.33\textwidth}
		\centering
		\includegraphics[width=\textwidth]{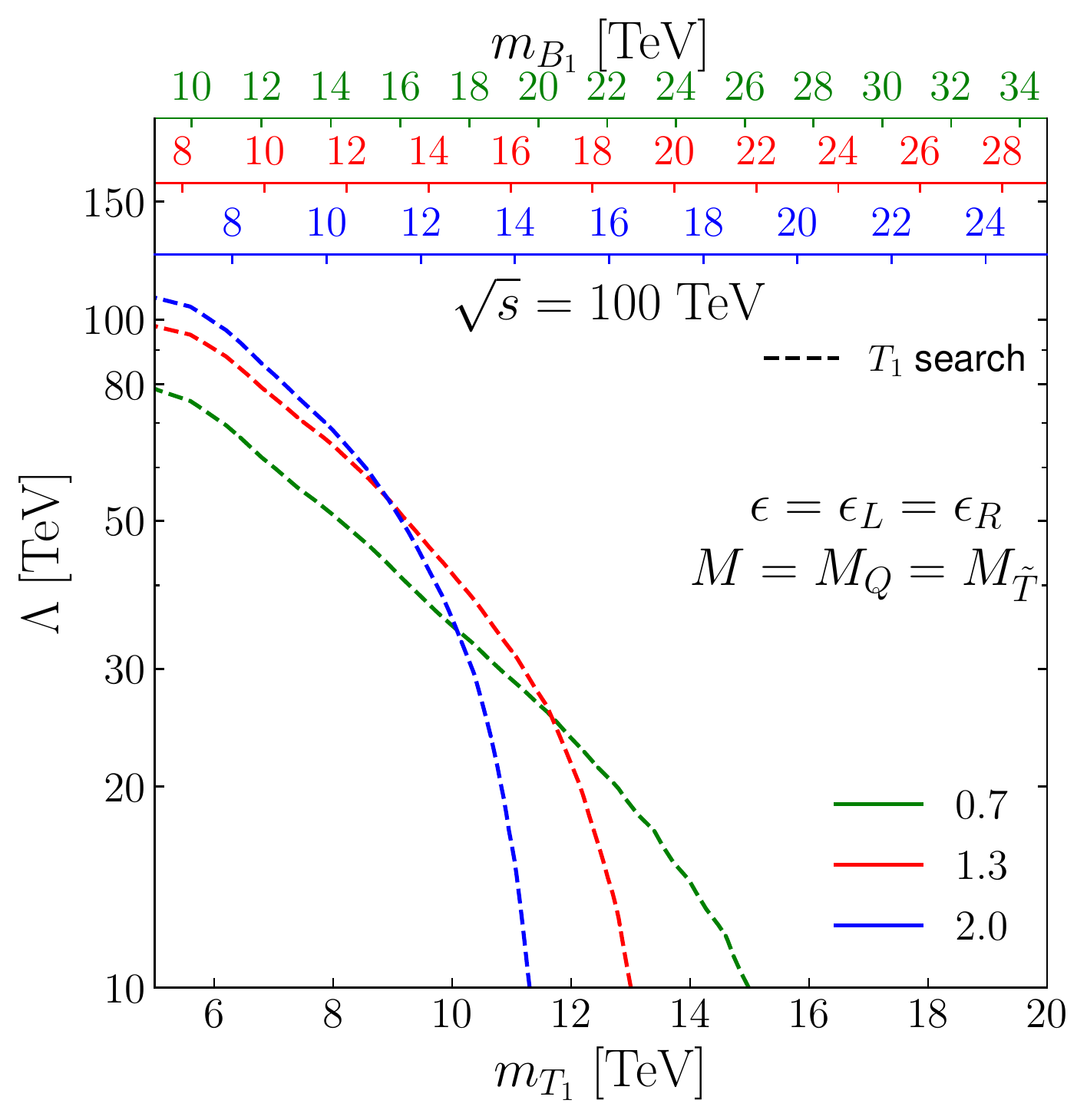}
		\caption{}
	\end{subfigure}
	\begin{subfigure}{0.33\textwidth}
		\centering
		\includegraphics[width=\textwidth]{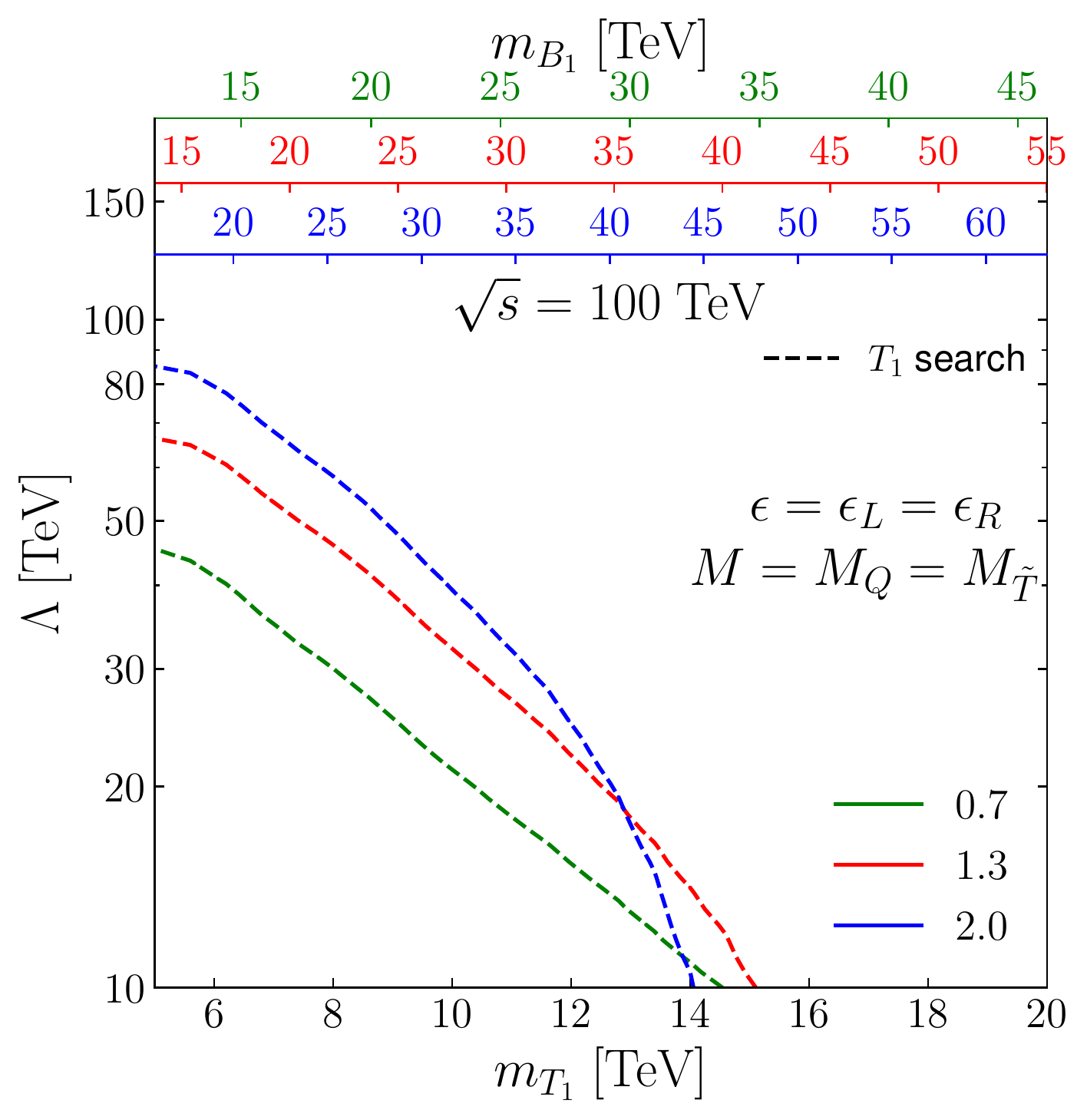}
		\caption{}
	\end{subfigure}
	\caption{Same as Fig.~\ref{figs:100TeV_final_1} but for $M_Q=2 M_{\tilde{T}}$.}
	\label{figs:100TeV_final_2}
\end{figure}

In this section, we estimate the sensitivity to our signal at a 100 TeV FCC-hh/SppC collider, with anticipated luminosity of $L = 30~{\rm fb}^{-1}$. We made use of the standard simulation chain described above, with a slight modification of the {\sc Delphes}~3 configuration shipped with the program. We used the default $b$-tagging efficiency formula
\begin{equation}
    \epsilon_b(p_T) = \epsilon_0(1-\frac{p_T}{Q})
\end{equation}
with $Q$ changed from 5~TeV (designed for the LHC)  to 50~TeV to have a realistic setup for 100 TeV collider.
 Moreover, we modified the event selection that can be found in~\cite{PhysRevD.104.095024} by requiring at least one slim $b$-jet, instead of two, that is contained in the Higgs fat jet. This is motivated by much higher boost featured by a Higgs boson (in comparison to the LHC) originating from a multi-TeV top partner decay, which makes it difficult to resolve the two $b$-jets. These modifications essentially improve  efficiencies  for the signal reconstruction and selection
 making them similar to those found in  the LHC studies.
 The results are displayed in Figs.~\ref{figs:100TeV_final_1} and \ref{figs:100TeV_final_2} for the same scenarios as studied in Section~\ref{sec:hllhc}.

We observe an important extension of the sensitivity as compared to the HL-LHC case, both in terms of vector-like quark masses reachable and the composite scale. However, this enhancement is smaller than what could be expected from a simple rescaling of the signal and background cross sections. This is due to the collimation of the decay products of the boosted heavy states, that spoils naive linear scaling. For light top partners of a few TeV composite scales of 80--100~TeV can then be reached, whereas the FCC-hh/SppC is sensitive to partners of about 10~TeV for lighter compositeness scale of about 10~TeV.

\section{Executive Summary}
\label{sec:executive-summary}

We studied the potential to extend searches for the vector-like partners of the third-generation Standard Model quarks on the basis of their expected chromomagnetic interactions. We first explored the region of the parameter space in which the bottom-quark partner is heavier than the top-quark partner, in which case the top-partner can be primarily produced via  the decay of the bottom-partner. Next, we probed the potential of the production of a single top-quark partner in association with an ordinary top-quark by gluon-fusion. 

We examined the sensitivity of these modes in the case where the top-partner subsequently decays to a Higgs boson and a SM top-quark, and demonstrated that these new channels have the potential of extending and complementing the conventional strategies at LHC run III and at the high-luminosity phase of the LHC. We found that partner masses that range up to about 3~TeV and 15--20 TeV can be reached at the HL-LHC and the FCC-hh/SppC colliders respectively, and that our analyses are correspondingly sensitive to composite scales of 30~TeV and 100~TeV for light top partners of a few TeV. This substantially expands the expected mass reach for these new states, including regions of parameter space that are inaccessible by traditional searches.

\section*{Acknowledgements}
The authors are grateful to Thomas Flacke for the organization of a focus meeting on Fundamental Composite Dynamics at IBS CTPU (Daejeon, Korea) in 2017, where the discussions that have given rise to this work have been initiated. 
Authors acknowledge the use of the IRIDIS High Performance Computing Facility, and associated support services
at the University of Southampton to complete this work. 
RSC, EHS, and XW were supported, in part, by by the US National Science Foundation under Grant No. PHY-1915147 and PHYS-2210177.
AB acknowledges partial  support from the STFC grant ST/L000296/1 and Soton-FAPESP grant.

\bibliographystyle{apsrev4-1.bst}

\bibliography{tstar}{}

\end{document}